\documentclass{article}
\usepackage{spconf,amsmath,graphicx}
\usepackage{amssymb}
\usepackage{algpseudocode}
\usepackage{cite}
\usepackage{amsfonts}
\usepackage{graphicx,subfigure}
\usepackage{fancyhdr}  %
\usepackage{cases}
\usepackage{extarrows}
\usepackage{algorithm}
\usepackage{multirow,tabularx}
\usepackage{mathtools}

\newtheorem{lemma}{Lemma}
\newtheorem{proposition}{Proposition}

\usepackage{color,xcolor}
\usepackage{etoolbox}
\usepackage{float}

\title{Achievable Rate maximization by Passive Intelligent Mirrors}
\name{Chongwen Huang$^{1}$, Alessio Zappone$^{2}$, M\'{e}rouane Debbah
$^{2,3}$, Chau Yuen$^{1}$ \thanks{The research of A. Zappone and M. Debbah is supported by the H2020 MSCA IF BESMART, Grant 749336. The work of M. Debbah is also funded by the H2020-ERC PoC-CacheMire, Grant 727682. The research of C. Huang and C. Yuen is supported A$^{\star}$STAR SERC project, Grant 1420200043 and  NSFC project, Grant 61750110529.}
}
\address{1: Singapore University of Technology and Design (SUTD), 8 Somapah Rd, 487372, Singapore\\
2: Large Networks and Systems Group (LANEAS), Laboratoire des Signaux et Systèmes\\  CentraleSupelec, CNRS, Univ. Paris Sud, Univ. Paris-Saclay, 91192 Gif-sur-Y vette, France \\
3: Mathematical and Algorithmic Sciences Laboratory, \\France Research Center, Huawei Technologies, Paris, France }
\begin{document}
%
\maketitle
\begin{abstract}
This paper investigates the use of a Passive Intelligent Mirrors (PIM) to operate a multi-user MISO downlink communication. The transmit powers and the mirror reflection coefficients are designed for sum-rate maximization subject to individual QoS guarantees to the mobile users. The resulting problem is non-convex, and is tackled by combining alternating maximization with the majorization-minimization method. Numerical results show the merits of the proposed approach, and in particular that the use of PIM increases the system throughput by at least $40\%$, without requiring any additional energy consumption.
\end{abstract}
\section{Introduction}\label{sec:intro}
The number of wireless devices is anticipated to reach $50$ billions by 2020 \cite{EricssonWP,scaling_up,hcw16,mmwave_5G,tan_indoor,hcw16_J}. This poses serious sustainability concerns for future cellular networks, which are demanded to provide 1000$\times$ higher data rates compared to present networks, while at the same time halving energy consumptions \cite{5GNGMN,Zap2016,Green_com}. This puts forth the critical needs for green and energy-efficient wireless solutions. The recent tutorial \cite{ZapNow15} and survey \cite{GEJSAC16} provide a comprehensive review of green solutions, including renewable energy sources, energy-efficient hardware, and green radio resource allocation and transmission.

One recent technology that has a significant potential in reducing energy consumptions in wireless networks is the so-called Passive Intelligent Mirrors (PIM), i.e. a physical meta-surface composed of many small-unit reflectors equipped with simple low-cost sensors and a cognitive engine. Each unit of the mirror is able to reflect a phase-shifted version of an incoming electromagnetic field. Thus, by suitably designing the phase shifts, it is possible to constructively combine the signals reflected by the different units, which  effectively makes the PIM an active medium. In other words, a PIM is able to act as an amplify-and-forward relay, but without requiring any dedicated energy source, thereby saving precious energy and enabling communication also in the presence of poor channel conditions. In addition, deployment costs are limited. A PIM can be easily integrated into the walls of the buildings surrounding the transmitter, as well as into ceilings of moving trains, laptop cases, and even on  people's arms.

Previous research on PIM mainly focused on indoor scenarios \cite{Subrt_control,tan_indoor,Subrt_control01,sha_hu,Reconfigurable_arrays}. In \cite{Subrt_control,Subrt_control01}, the concept of intelligent wall was proposed, as a wall that can be  equipped inside a building to improve indoor communications. In \cite{sha_hu}, a detailed analysis on the information transfer from the users to the large intelligent surfaces was carried out. However, these previous works did not provide any system design method, focusing merely on introducing the PIM idea in indoor scenarios.

In contrast, this work focuses on an outdoor scenario, considering a MISO downlink channel, and optimizing the transmit powers and the PIM phase shifts so as to maximize the system sum-rate. The resulting optimization problem is non-convex, and a provably convergent, low-complexity optimization method is developed merging the alternating optimization and majorization-minimization frameworks. Numerical results show that PIM can increase the system sum-rate by more than $40\%$, without any additional energy consumption.

\section{System Model}\label{sec:format}
\begin{figure}
  \begin{center}
  \includegraphics[width=74mm]{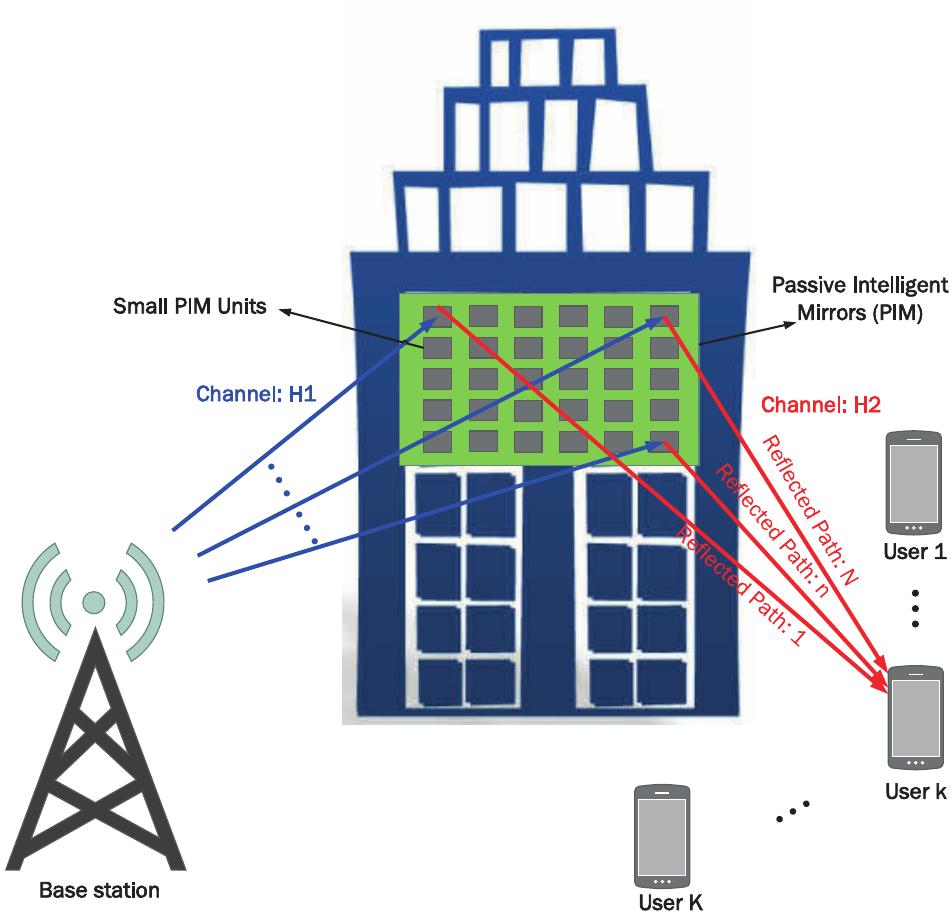}  \vspace{-4mm} %
  \caption{A BS serves $K$ mobile users through a PIM}
  \label{fig:Estimation_Scheme} \vspace{-6mm}
  \end{center}
\end{figure}
The considered system model is shown in Fig. \ref{fig:Estimation_Scheme}. A base station (BS) equipped with $M$ antennas serves $K$ single-antenna users, through a PIM composed of $N$ reflecting units embedded in a surrounding building, which acts a relay. The direct path between the BS and the mobile is neglected as it is assumed that no line-of-sight communication is present.
Then, the discrete-time signal received at user $k$ can be written as
\begin{equation}\label{model_01}
  y_k=  \mathbf{h}_{k,2}\mathbf{\Theta} \mathbf{H}_{1}\mathbf{x}+w_k, k=1,2,..., K,
\end{equation}
where $\mathbf{h}_{k,2}\in \mathbb{C}^{1\times N}$ and $\mathbf{H}_{1}\in \mathbb{C}^{N\times M}$ denote the channel between the PIM and user $k$ and between the BS and PIM, with entries modeled as independent and identically distributed (i.i.d.) complex circularly symmetric standard Gaussian variables; $\mathbf{\Theta}=\mathrm{diag}[e^{j\theta_{1}},\ldots,e^{j\theta_{N}}]$ is the PIM phase-shift matrix, with $j$ the imaginary unit; $w_k \sim \mathcal{CN} (0, \sigma^2)$ models the thermal noise at user $k$; $\mathbf{x}=\sum_{k=1}^{K}\sqrt{p}_{k}\mathbf{g}_{k}s_{k}$ is the BS transmit signal, with $p_{k}$, $s_{k}$, and $\mathbf{g}_{k}$ the transmit power, information symbol, and beamforming vector intended for user $k$.
Then, for all $k$, the SINR enjoyed by user $k$ is
\begin{equation}\label{model_05}
  \gamma_k =\frac{p_k|\mathbf{h}_{2,k}\mathbf{\Theta} \mathbf{H}_{1}\mathbf{g}_k|^2}{\sum\limits_{i=1,i\neq k}^{K} p_i|\mathbf{h}_{2,k}\mathbf{\Theta} \mathbf{H}_{1}\mathbf{g}_i|^2+\sigma^2}.
\end{equation}
\subsection{Problem formulation}
The goal of this work is to optimize the transmit powers and the matrix $\mathbf{\Theta}$ for system sum-rate maximization. To make the problem tractable, we employ zero-forcing transmission, which is optimal in the high-SNR regime \cite{DE_ZFR,de_MIMO}. Assuming $N\geq K$, and defining $\mathbf{G}=[\mathbf{g}_1,\ldots,\mathbf{g}_K] \in \mathbb{C}^{M\times K}$ and $\mathbf{H}_2=[\mathbf{h}_{2,1}^{T},\ldots,\mathbf{h}_{2,K}^{T}]^{T}\in \mathbb{C}^{K\times N} $, zero-forcing is achieved by setting $\mathbf{G}=(\mathbf{H}_{2}\mathbf{\Theta} \mathbf{H}_1)^{+}$, with $(\cdot)^{+}$ denoting pseudo-inversion. Then, defining the matrix $\mathbf{P}=\text{diag}[p_{1},\ldots,p_{K}]$ and denoting the maximum feasible transmit power at the BS by $P_{max}$, and the $k$-th user's rate requirement by $R_{min,k}$, the sum-rate maximization problem can be formulated as
\begin{subequations}\label{Prob:ResAllProb}
\begin{align}
&\displaystyle \max_{\mathbf{\Theta},\mathbf{P}}\;\sum_{k=1}^{K}\log_2\left(1+\frac{p_k}{\sigma^2}\right)\label{Prob:aResAllProb}\\
&\;\text{s.t.}\;\log_2\left(1+\frac{p_k}{\sigma^2}\right)\geq R_{min,k}\;,\forall\;k=1,\ldots,K\label{Prob:bResAllProb}\\
&\;\quad\;\;\text{tr}((\mathbf{H}_{2}\mathbf{\Theta} \mathbf{H}_{1})^{+}\mathbf{P}(\mathbf{H}_{2}\mathbf{\Theta} \mathbf{H}_{1})^{+H})\leq P_{max}\label{Prob:cResAllProb}\\
&\;\quad\;\;0\leq \theta_{i}\leq 2\pi\;,\forall\;i=1,\ldots,N\label{Prob:dResAllProb}
\end{align}
\end{subequations}
Problem \eqref{Prob:ResAllProb} is non-convex, and especially the optimization with respect to $\mathbf{\Theta}$ appears challenging \cite{mudulors_stoica}. The next section introduces a computationally-affordable way to tackle \eqref{Prob:ResAllProb}.

\section{Proposed approach}
In order to tackle \eqref{Prob:ResAllProb} with affordable complexity, a convenient approach is to separately and iteratively optimize $\mathbf{P}$ and $\mathbf{\Theta}$.

\subsection{Optimization with respect to $\mathbf{\Theta}$}
For fixed $\mathbf{P}$, Problem \eqref{Prob:ResAllProb} becomes the feasibility test
\begin{subequations}\label{Eq:ResAllProbTheta}
\begin{align}
&\displaystyle \max_{\mathbf{\Theta}}\;1\\
&\;\text{s.t.}\;\text{tr}((\mathbf{H}_{2}\mathbf{\Theta} \mathbf{H}_{1})^{+}\mathbf{P}(\mathbf{H}_{2}\mathbf{\Theta} \mathbf{H}_{1})^{+H})\leq P_{max}\label{Eq:bResAllProbTheta}\\
&\;\quad\;\;0\leq \theta_{i}\leq 2\pi\;,\forall\;i=1,\ldots,N\label{Eq:cResAllProbTheta}
\end{align}
\end{subequations}
As a first step, it is convenient to apply the change of variable $\phi_{k}=e^{j\theta_{k}}$, which leads to the problem:
\begin{subequations}\label{Eq:ResAllProbPhi}
\begin{align}
&\displaystyle \max_{\mathbf{\mathbf{\Phi}}}\;1\\
&\;\text{s.t.}\;\text{tr}((\mathbf{H}_{2}\mathbf{\Phi} \mathbf{H}_{1})^{+}\mathbf{P}(\mathbf{H}_{2}\mathbf{\Phi} \mathbf{H}_{1})^{+H})\leq P_{max}\label{Eq:bResAllProbPhi}\\
&\;\quad\;\;|\phi_{i}|= 1,\forall\;i=1,\ldots,N\;,\label{Eq:cResAllProbPhi}
\end{align}
\end{subequations}
The challenge in solving Problem \eqref{Eq:ResAllProbPhi} lies in the fact that its objective is non-differentiable and that \eqref{Eq:cResAllProbPhi} is a non-convex constraint. To proceed further, we observe that \eqref{Eq:ResAllProbPhi} is feasible if and only if the solution of the problem
\begin{subequations}\label{Eq:ResAllProbPhi2}
\begin{align}
&\displaystyle \min_{\mathbf{\mathbf{\Phi}}}\;\text{tr}((\mathbf{H}_{2}\mathbf{\Phi}\mathbf{H}_{1})^{+}\mathbf{P}(\mathbf{H}_{2}\mathbf{\Phi} \mathbf{H}_{1})^{+H})\label{Eq:aResAllProbPhi2}\\
&\;\text{s.t.}\;|\phi_{i}|= 1,\forall\;i=1,\ldots,N\;,\label{Eq:bResAllProbPhi2}
\end{align}
\end{subequations}
is such that the objective can be made lower than $P_{max}$.
At this point, we observe that \eqref{Eq:aResAllProbPhi2} can be rewritten as follows:
\begin{align}\label{Eq:ObjResAllProbPhi2}
&\text{tr}((\mathbf{H}_{2}\mathbf{\Phi}\mathbf{H}_{1})^{+}\mathbf{P}(\mathbf{H}_{2}\mathbf{\Phi} \mathbf{H}_{1})^{+H})=\|\mathbf{H}_{1}^{+}\mathbf{\Phi}^{-1}\mathbf{H}_{2}^{+}\|_{F}\notag\\
=&\|\text{vec}(\mathbf{H}_{1}^{+}\mathbf{\Phi}^{-1}\mathbf{H}_{2}^{+})\|^{2}=\|(\mathbf{H}_{2}^{+}\otimes\mathbf{H}_{1}^{+})\text{vec}(\mathbf{\Phi}^{-1})\|^2\;,
\end{align}
where we have exploited the properties of the Frobenius matrix norm and the connection
between the vectorization operator and the Kronecker product.
As we will show, the objective in \eqref{Eq:ObjResAllProbPhi2} enables to deal with the non-convex constraint \eqref{Eq:bResAllProbPhi2}, provided one is able to reformulate \eqref{Eq:ObjResAllProbPhi2} into a differentiable function. To this end, a convenient approach is to resort to the Majorization-Minimization (MM) method \cite{MM_review,MM_sun,MM_song}. The MM method is an iterative approach that, in the $i$-th iteration, maximizes an upper-bound of \eqref{Eq:ObjResAllProbPhi2}. However, for any $i$-th iteration, the upper-bound maximized in $i$-th iteration  and the true objective \eqref{Eq:ObjResAllProbPhi2} must be equal when evaluated at the maximizer computed in $(i-1)$-th iteration. The MM is attractive because it enjoys the monotonic improvement property, i.e., it monotonically decreases the value of the true objective \eqref{Eq:ObjResAllProbPhi2} after each iteration. This also implies converges in the objective value, since \eqref{Eq:ObjResAllProbPhi2} is lower-bounded over the problem feasible set. Nevertheless, the challenge in using the MM method lies in determining a suitable upper bound of \eqref{Eq:ObjResAllProbPhi2}, which fulfills the theoretical requirements of the method, (i.e., coincides with the true objective at a given point), and is also easier to minimize compared with the original objective. For the case at hand, a convenient upper-bound is provided in the following lemma.
\begin{lemma}\label{Lem:1}
Consider \eqref{Eq:ResAllProbPhi2}. Then, for any feasible $\mathbf{x}=\mathrm{vec} (\mathbf{\Phi}^{-1})$, and given any feasible point $\mathbf{x_{t}}$, a suitable upper-bound to employ the MM method is:
\begin{align}\label{eq_5}
&\|(\mathbf{H}_{2}^{+}\otimes\mathbf{H}_{1}^{+} )\mathbf{x}\|^{2}\leq f(\mathbf{x}|\mathbf{x}_t)=\notag\\
&\mathbf{x}^H\mathbf{M}\mathbf{x}+2\mathrm{Re}(\mathbf{x}^H(\mathbf{L}-\mathbf{M})\mathbf{x_t})+\mathbf{x_t}^H(\mathbf{M}-\mathbf{L})\mathbf{x_t}^H,
\end{align}
wherein, $\mathbf{M}=c_t^{max}\mathbf{A}\mathbf{A}^H$, $\mathbf{L}= \mathbf{A}\left( \mathrm{diag}(\mathbf{c}_t)-N^2\mathbf{I}\right)\mathbf{A}^H$,
$\mathbf{A}=\mathbf{H}_{2}^{+}\otimes\mathbf{H}_{1}^{+}$, $\mathbf{c}_t = |\mathbf{A^Hx_t}|$, and $c_t^{max}=\mathrm{max}(\mathbf{c}_t)$ is a maximum element in the vector $\mathbf{c}_t$.
\end{lemma}

\noindent {\bf Proof:}
The proof leverages the second-order Taylor expansion of \eqref{eq_5}. Full details are omitted due to space constraints.

Based on Lemma \ref{Lem:1}, in each iteration of the MM method we are faced by the problem
\begin{subequations}\label{Eq:theta3}
\begin{align}
& \min_{\mathbf{x}}\;  f(\mathbf{x}|\mathbf{x}_t)\label{Eq:theta3a}\\
&\; \text {s.t. } \; |x_{i}|=1\;,\forall,i=1,\ldots,N\;.\label{Eq:theta3b}
\end{align}
\end{subequations}
\begin{proposition}
For any $\mathbf{x}_t$, \eqref{Eq:theta3} is solved by $\mathbf{x}=e^{-j\mathrm{arg}(\mathbf{y})}$, with $\mathrm{arg}(\mathbf{y})$ denoting the component-wise phase of $\mathbf{y}$, and
\begin{equation}\label{eq_12}
\mathbf{y}=\frac{-(\mathbf{A}\left( \mathrm{diag}(\mathbf{c}_t)-c_t^{max}\mathbf{I}-N^2\mathbf{I}\right)\mathbf{A}^H)}{c_t^{max}\mathbf{A}\mathbf{A}^H}\mathbf{x_t}
\end{equation}
\end{proposition}
{\bf Proof:}
The result follows from the analysis of the stationary points of \eqref{Eq:theta3a}. Details are omitted due to space constraints.

\subsection{Optimization with respect to $\mathbf{P}$}\label{Sec:PowerAll}
For fixed $\mathbf{\Theta}$, Problem \eqref{Prob:ResAllProb} is stated as,
\begin{subequations}\label{Eq:theta4}
\begin{align}
&\displaystyle
\max_{\mathbf{P}}\;\sum_{k=1}^{K}\log_2\left(1+\frac{p_k}{\sigma^2}\right)\label{Eq:theta41}\\
&\; \text {s.t. } \;p_{k}\geq \sigma^{2}(2^{R_{min,k}}-1)\;,\forall\;k=1,\ldots,K \label{Eq:theta42}\\
&\;\;\quad\;\;\text{tr}((\mathbf{H}_{2}\mathbf{\Theta} \mathbf{H}_{1})^{+}\mathbf{P}(\mathbf{H}_{2}\mathbf{\Theta} \mathbf{H}_{1})^{+ H})\leq P_{max}\label{Eq:theta43}
\end{align}
\end{subequations}
Problem \eqref{Eq:theta4} is convex and thus can be solved by means of standard convex optimization techniques \cite{yuwei,Jindal}. Specifically, analyzing the Karush--Kuhn--Tucker (KKT) optimality conditions of \eqref{Eq:theta4} yields the following closed-form expression for the solution of \eqref{Eq:theta4}.
\begin{lemma}\label{Lem:2}
Problem \eqref{Eq:theta4} admits the following solution
\begin{equation}\label{eq_13}
p_k=[\alpha\lambda_k-\sigma^2]^{+}+\sigma^{2}(2^{R_{min,k}}-1)\lambda_k^{-1},
\end{equation}
where water level $\alpha=\frac{1}{q}(P_{max}-\sum_{k=1}^K\sigma^{2}(2^{R_{min,k}}-1)\lambda_k^{-1}+ \sigma^2\sum_{k=1}^q\lambda_k^{-1})$ is the Lagrange multiplier associated to \eqref{Eq:theta43},
$\lambda_k$ is the $k$-th eigenvalue of $(\mathbf{H}_{2}\mathbf{\Theta} \mathbf{H}_{1})$ $(\mathbf{H}_{2}\mathbf{\Theta} \mathbf{H}_{1})^{H}$, $q$ is the number of non-zero eigenvalues $\lambda_{k}$, and $[x]^+$  denotes max (0, x).  
\end{lemma}
Finally, the overall algorithm can be stated as in Algorithm 1, wherein the inner loop implements the MM algorithm, and the outer loop the alternating maximization. We remark that each iteration of Algorithm \ref{euclid} monotonically increases the value of the overall problem objective in \eqref{Prob:aResAllProb}. Thus, Algorithm 1 converges in the value of the objective, since \eqref{Prob:aResAllProb} is continuous over the compact feasible set of \eqref{Prob:ResAllProb}, and thus is upper-bounded.
\begin{algorithm}
\caption{ Proposed Sum-rate Maximization Algorithm}\label{euclid}
\begin{algorithmic}[1]
\State \textbf{Require}: $P_{max}$, $\sigma^2$, $\{R_{min,k}\}_{k=1}^{K}$, $\mathbf{H}_{2}$, and $\mathbf{H}_{1}$;
\State \textbf{Initialize} $\{p_{k}\}_{k=1}^{K}$, and set $t$=0;
\State \textbf{Repeat}:
\State \quad\textbf{Repeat}:
\State \qquad $\mathbf{c}_t = |\mathbf{A^Hx_t}|$; $c_t^{max}=\mathrm{max}(\mathbf{c}_t)$; $\mathbf{M}=c_t^{max}\mathbf{A}\mathbf{A}^H$;
\State\qquad $\mathbf{L}=\mathbf{A}\left(\mathrm{diag}(\mathbf{c}_t)-N^2\mathbf{I}\right)\mathbf{A}^H$;
\State \qquad $\mathbf{y}=\frac{-(\mathbf{A}\left( \mathrm{diag}(\mathbf{c}_t)-c_t^{max}\mathbf{I}-N^2\mathbf{I}\right)\mathbf{A}^H)}{c_t^{max}\mathbf{A}\mathbf{A}^H}\mathbf{x_t}$;
\State \qquad $\mathbf{x}_{t+1} =e^{j\mathrm{arg}(\mathbf{y})}$; $t \leftarrow t+1$;
\State \quad \textbf{Until} Convergence is reached; Obtain $\mathbf{\Phi}$;
\State \qquad\textbf{If} \eqref{Eq:aResAllProbPhi2} evaluated at $\mathbf{\Phi}$ is lower than $P_{max}$:
\State \qquad $p_k=[\alpha\lambda_k-\sigma^2]^{+}+\sigma^{2}(2^{R_{min,k}}-1)\lambda_k^{-1}$, for all $k$;
\State \qquad\textbf{Else}
\State \qquad\quad Break and declare unfeasibility.
\State \textbf{Until} Convergence is reached; $\mathbf{\Phi}$ and $p_k$ are obtained.
\end{algorithmic}
\end{algorithm}


\section{NUMERICAL RESULTS}
In our numerical simulations, we considered the system scenario described in Section \ref{sec:format}, with users randomly placed in an area of 625 $m^2$ with channels generated as realizations of random matrices with i.i.d. entries drawn from a standard complex Gaussian distribution. All results have been obtained by averaging over 500 independent channels and users positions realizations. We define the Signal Noise Ratio (SNR), SNR=$P_{max}/\sigma^2$.
\begin{figure}[t]
  \begin{center}
  \includegraphics[width=80mm]{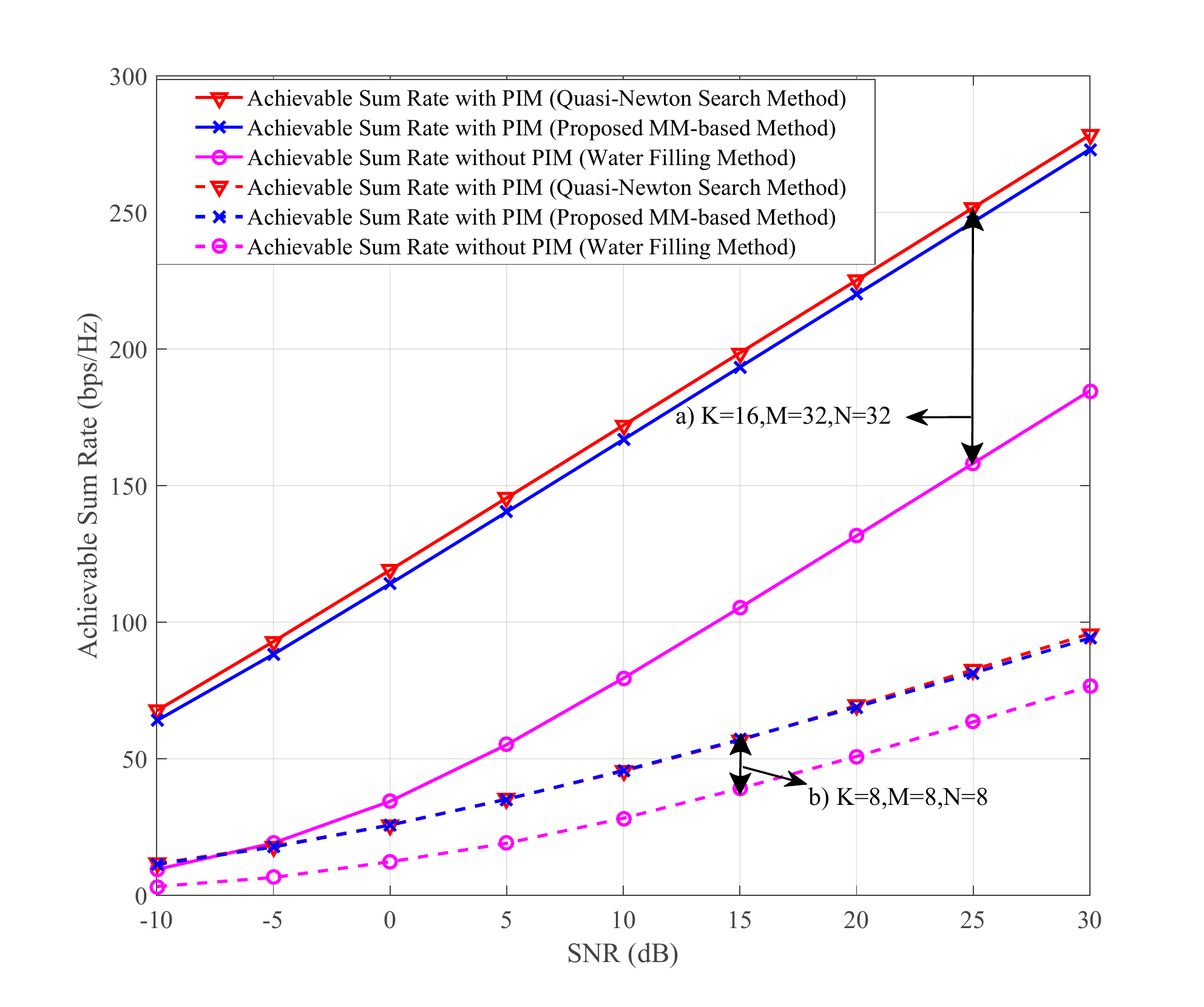}  \vspace{-2mm}
  \caption{Achievable rate versus SNR. $R_{min}=\log_2(1+\frac{\mathrm{SNR}}{2K})$ bps/Hz; a) $K=16$, $M=32$, $N=32$; b) $K=8$, $M=8$, $N=8$.}
  \label{fig:ombing} \vspace{-0mm}
  \end{center}
\end{figure}
\begin{figure}[t]
  \begin{center}
  \includegraphics[width=80mm]{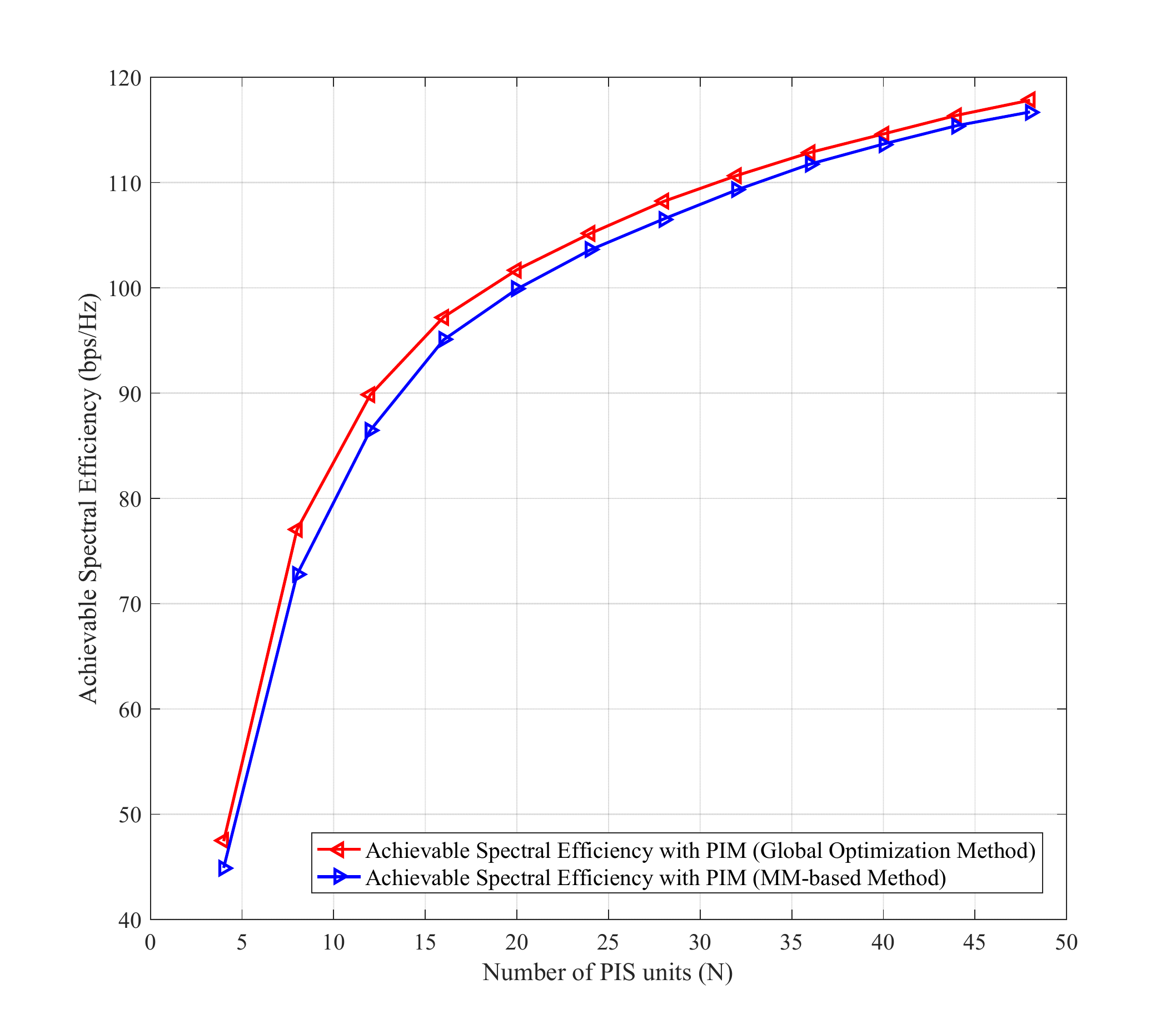}  \vspace{-2mm} %
  \caption{Spectral Efficiency versus PIM units number. SNR$=20$ dB, $K=16$, $M=8$, $R_{min,k}=2$ bps/Hz: a) Global optimization method; b) Proposed MM-based method.}
  \label{fig:N_8_40} \vspace{-1mm}
  \end{center}
\end{figure}
Fig. \ref{fig:ombing} compares the achievable sum rate versus SNR. We consider two sets of system parameters, namely $K=16$, $M=32$, $N=32$ and $K=8$, $M=8$, $N=8$, and the minimum QoS rate has been set to $R_{min,k}=R_{min}= \log_2(1+\frac{\mathrm{SNR}}{2K})$ bps/Hz for all $k=1,\ldots,K$ according to the different SNRs. The optimal solution of problem \eqref{Prob:ResAllProb} is obtained through global optimization methods (Quasi-Newton search). This approach has an exponential complexity and is considered here only for benchmarking purpose. Resource allocation in a system without PIM is also considered as a baseline scheme. In this case only power allocation needs to be performed, which is accomplished by the tools described in Section 3.2. We compare the achievable sum rate of the proposed Algorithm \ref{euclid} with the optimal solution. It can be seen that employing the PIM increases the sum rate by more than $40\%$. Also, the gap becomes wider as the number of antennas and PIM units increases. Furthermore, Algorithm \ref{euclid} suffers a limited gap compared to the global optimization method, which has a much higher complexity.
\label{sec:majhead}

Fig. \ref{fig:N_8_40} shows the achievable sum spectral efficiency of the global optimization method and proposed Algorithm \ref{euclid}, versus the number of PIM units. The SNR is $20\,\textrm{dB}$, while $K=16$, $M=8$, $R_{min,k}=2$ bps/Hz for all $k$. It can be seen that the proposed Algorithm \ref{euclid} matches the sum spectral efficiency obtained by the global optimization method. Furthermore, this figure also shows that the more PIM units the larger the sum spectral efficiency in one cell, even though, as expected, the increase saturates as the number of PIM units grows.

\begin{figure} \vspace{-15mm}
  \begin{center}
  \includegraphics[width=80mm]{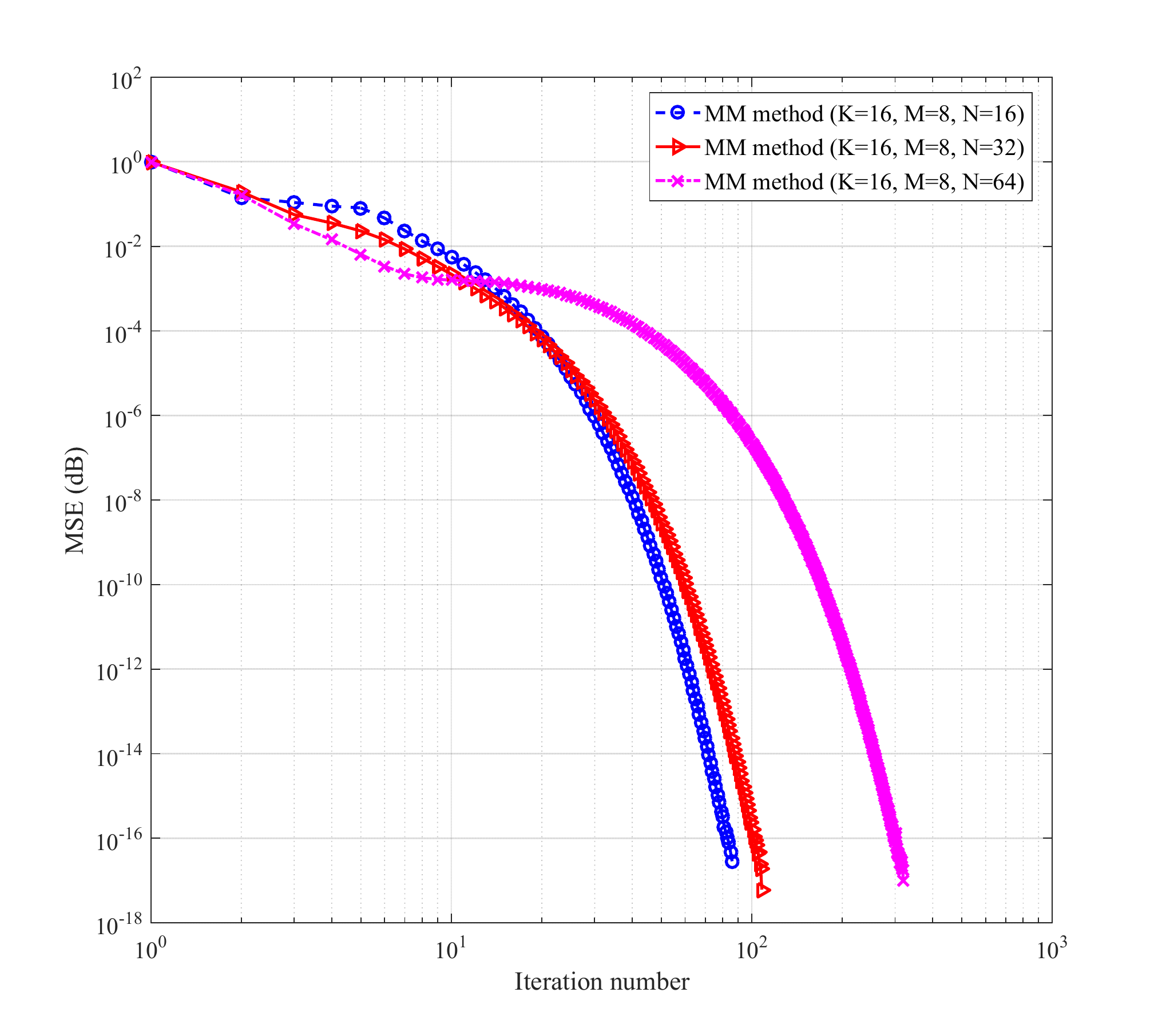}  \vspace{-2mm} %
  \caption{ MSE versus iteration number. $K=16$, $M=8$, $N=16;32;64$.}
  \label{fig:MM_iteration} \vspace{-3mm}
  \end{center}
\end{figure}

Finally, Fig. \ref{fig:MM_iteration} addresses the convergence speed of the MM-based method employed within Algorithm \ref{euclid}, in terms of the number of iterations to reach a given Mean Square Error (MSE) among two successive iterations, defined as
\begin{equation}
\mathrm{MSE}= \frac{\|\mathbf{\Theta}_{t+1}-\mathbf{\Theta}_{t}\|^{2}}{\|\mathbf{\Theta}_{t}\|^{2}}
\end{equation}
System parameters have been set to $K=16$, $M=8$, and $N=16;32;64$. It is seen that Algorithm \ref{euclid} reaches acceptable values of MSE in a few dozens of iterations, which increases as $N$ grows larger, since this corresponds to increasing the number of optimization variables. Nevertheless, recalling that each iteration of the MM method involves simple closed-form computations, Fig. \ref{fig:MM_iteration} confirms the very limited complexity of the proposed MM-based method.

\section{Conclusion}\label{sec:prior}
A sum-rate maximization scheme for a  PIM-based, multi-user MIMO system was developed. The non-convex radio resource allocation problem was tackled by combining MM and alternating optimization, to provide a provably convergent and low-complexity algorithm. Numerical result show that the proposed scheme achieves near-optimal performance, and improves by more than $40\%$ the sum rate compared to traditional systems without PIM.

\clearpage

\bibliographystyle{IEEEbib}
\bibliography{strings,refs}

\begin{thebibliography}{10}

\bibitem{EricssonWP}
{Ericsson White Paper},
\newblock ``More than 50 billion connected devices,''
\newblock Tech. {R}ep. 284 23-3149 Uen, Ericsson, Feb. 2011.

\bibitem{scaling_up}
F.~Rusek, D.~Persson, B.~K. Lau, E.~G. Larsson, T.~L.Marzetta, O.~Edfors, and
  F.~Tufvesson,
\newblock ``Scaling up mimo: Opportunities and challenges with very large
  arrays,''
\newblock {\em IEEE Signal Processing Magazine}, vol. 30, no. 1, pp. 40--60,
  2013.

\bibitem{hcw16}
C.~Huang, L.~Liu, C.~Yuen, and S.~Sun,
\newblock ``A lse and sparse message passing-based channel estimation for
  mmwave mimo systems,''
\newblock in {\em 2016 IEEE Globecom Workshops (GC Wkshps)}, Dec 2016, pp.
  1--6.

\bibitem{mmwave_5G}
T.~S. Rappaport, S.~Sun, R.~Mayzus, Y.~Azar H.~Zhao, K.~Wang, and F.~Gutierrez,
\newblock ``Millimeter wave mobile communications for 5{G} cellular: It will
  work!,''
\newblock {\em IEEE access}, vol. 1, pp. 335--349, 2013.

\bibitem{tan_indoor}
X.~Tan, Z.~Sun, J.~M. Jornet, and D.~Pados,
\newblock ``Increasing indoor spectrum sharing capacity using smart
  reflect-array,''
\newblock in {\em 2016 IEEE International Conference on Communications (ICC)},
  May 2016, pp. 1--6.

\bibitem{hcw16_J}
C.~Huang, L.~Liu, C.~Yuen, and S.~Sun,
\newblock ``Iterative channel estimation using lse and sparse message passing
  for mmwave mimo systems,''
\newblock {\em arXiv preprint arXiv:1611.05653}, 2016.

\bibitem{5GNGMN}
``{NGMN} alliance 5{G} white paper,''
\newblock {\em https://www.ngmn.org/5g-white-paper/5g-white-paper.html}, 2015.

\bibitem{Zap2016}
A.~Zappone, L.~Sanguinetti, G.~Bacci, E.~Jorswieck, and M.~Debbah,
\newblock ``Energy-efficient power control: A look at 5{G} wireless
  technologies,''
\newblock {\em IEEE Transactions on Signal Processing}, vol. 64, no. 7, pp.
  1668--1683, April 2016.

\bibitem{Green_com}
Y.~Chen, S.~Zhang, S.~Xu, and G.~Y. Li,
\newblock ``Fundamental trade-offs on green wireless networks,''
\newblock {\em IEEE Communications Magazine}, vol. 49, no. 6, pp. 30--37, June
  2011.

\bibitem{ZapNow15}
A.~Zappone and E.~A. Jorswieck,
\newblock ``Energy efficiency in wireless networks via fractional programming
  theory,''
\newblock {\em Foundations and Trends{\textregistered} in Communications and
  Information Theory}, vol. 11, no. 3-4, pp. 185--396, 2015.

\bibitem{GEJSAC16}
S.~Buzzi, C.-L. I, T.~E. Klein, H.~V. Poor, C.~Yang, and A.~Zappone,
\newblock ``A survey of energy-efficient techniques for 5{G} networks and
  challenges ahead,''
\newblock {\em IEEE Journal on Selected Areas in Communications}, vol. 34, no.
  5, 2016.

\bibitem{Subrt_control}
L.~Subrt and P.~Pechac,
\newblock ``Intelligent walls as autonomous parts of smart indoor
  environments,''
\newblock {\em IET Communications}, vol. 6, no. 8, pp. 1004--1010, May 2012.

\bibitem{Subrt_control01}
L.~Subrt and P.~Pechac,
\newblock ``Controlling propagation environments using intelligent walls,''
\newblock in {\em 2012 6th European Conference on Antennas and Propagation
  (EUCAP)}, March 2012, pp. 1--5.

\bibitem{sha_hu}
S.~Hu, F.~Rusek, and O.~Edfors,
\newblock ``The potential of using large antenna arrays on intelligent
  surfaces,''
\newblock {\em CoRR}, vol. abs/1702.03128, 2017.

\bibitem{Reconfigurable_arrays}
S.~V. Hum and J.~Perruisseau-Carrier,
\newblock ``Reconfigurable reflectarrays and array lenses for dynamic antenna
  beam control: A review,''
\newblock {\em IEEE Transactions on Antennas and Propagation}, vol. 62, no. 1,
  pp. 183--198, Jan 2014.

\bibitem{DE_ZFR}
S.~Wagner, R.~Couillet, M.~Debbah, and D.~T. Slock,
\newblock ``Large system analysis of linear precoding in correlated {MISO}
  broadcast channels under limited feedback,''
\newblock {\em IEEE Transactions on Information Theory}, vol. 58, no. 7, pp.
  4509--4537, 2012.

\bibitem{de_MIMO}
M.~Debbah and R.~R. Muller,
\newblock ``{MIMO} channel modeling and the principle of maximum entropy,''
\newblock {\em IEEE Transactions on Information Theory}, vol. 51, no. 5, pp.
  1667--1690, 2005.

\bibitem{mudulors_stoica}
P.~Stoica, H.~He, and J.~Li,
\newblock ``New algorithms for designing unimodular sequences with good
  correlation properties,''
\newblock {\em IEEE Transactions on Signal Processing}, vol. 57, no. 4, pp.
  1415--1425, April 2009.

\bibitem{MM_review}
D.~R. Hunter and K.~Lange,
\newblock ``A tutorial on {MM} algorithms,''
\newblock {\em The American Statistician}, vol. 58, no. 1, pp. 30--37, 2004.

\bibitem{MM_sun}
Y.~Sun, P.~Babu, and D.~P. Palomar,
\newblock ``Majorization-{M}inimization algorithms in signal processing,
  communications, and machine learning,''
\newblock {\em IEEE Transactions on Signal Processing}, vol. 65, no. 3, pp.
  794--816, Feb 2017.

\bibitem{MM_song}
J.~Song, P.~Babu, and D.~P. Palomar,
\newblock ``Optimization methods for designing sequences with low
  autocorrelation sidelobes,''
\newblock {\em IEEE Transactions on Signal Processing}, vol. 63, no. 15, pp.
  3998--4009, Aug 2015.

\bibitem{yuwei}
W.~Yu, W.~Rhee, S.~Boyd, and J.~M. Cioffi,
\newblock ``Iterative water-filling for {G}aussian vector multiple-access
  channels,''
\newblock {\em IEEE Transactions on Information Theory}, vol. 50, no. 1, pp.
  145--152, Jan 2004.

\bibitem{Jindal}
N.~Jindal, W.~Rhee, S.~Vishwanath, S.~A. Jafar, and A.~Goldsmith,
\newblock ``Sum power iterative water-filling for multi-antenna {G}aussian
  broadcast channels,''
\newblock {\em IEEE Transactions on Information Theory}, vol. 51, no. 4, pp.
  1570--1580, April 2005.

\end{thebibliography}

\end{document}